\documentclass[fleqn,twoside]{article}
\usepackage{espcrc2}
\usepackage{psfrag,epsfig,axodraw,amsmath,feynarts}
\usepackage[dvips]{color}
\usepackage{colordvi}

\title{Supersymmetric two-loop contributions to the anomalous magnetic
  moment of the muon}

\author{Dominik St\"ockinger\address{
Institute for Particle Physics 
Phenomenology, University of Durham,\\
Durham DH1~3LE, UK}\thanks{email: Dominik.Stockinger@durham.ac.uk}}

\input paperdef

\begin{document}

\begin{abstract}

Recent results of two interesting classes of supersymmetric two-loop
contributions to $(g-2)_\mu$ are presented. Two-loop diagrams
involving a closed sfermion loop can amount to $5\times10^{-10}$ or
almost one standard deviation of the current experimental
uncertainty, and two-loop diagrams with a closed chargino/neutralino
loop can be similarly large. The dependence of these two classes on
the unknown supersymmetric 
parameters and the sensitivity on existing experimental
constraints on the parameter space are quite different. 
We also comment on the calculational techniques, in particular the large
mass expansion and the treatment of $\ga_5$ and the $\epsilon$-tensor.

\end{abstract}

\maketitle

\section{Introduction}

In the past few years, the research on the anomalous magnetic moment
$\amu=(g-2)_\mu/2$ of the muon has made exciting progress. The
Brookhaven ``Muon $g-2$ Experiment'' (E821) has measured 
$\amu$ with unprecedented precision to $\amuexp = (11\, 659\, 208 \pm
6) \times 10^{-10}$ \cite{g-2exp}. Motivated by this tremendous
experimental precision, many groups have contributed to an improvement
of the evaluation of the hadronic contributions to $\amu$, the
bottleneck of the Standard Model (SM) prediction
\cite{DEHZ,g-2HMNT,Jegerlehner,Yndurain,LBL,LBLnew}. Now the
theoretical precision is at the same level as the experimental one. 

The comparison of the experimental value and the SM theory prediction
exhibits a discrepancy of about $(20\ldots30)\times10^{-10}$,
depending on the evaluation of the hadronic contributions. For the
purpose of the present talk we will use the recent evaluations of
\cite{g-2HMNT,LBLnew}, resulting in\footnote{This evaluation is
  $e^+e^-$ data driven.  
Recent analyses concerning $\tau$ data indicate that uncertainties due to
isospin breaking effects may have been underestimated
earlier~\cite{Jegerlehner}.}
\BEA
\label{deviationfinal}
\amuexp-\amutheo & = &
(24.5\pm9)\times10^{-10}.
\EEA
The precision is high enough that SM electroweak one- and two-loop
contributions, amounting to $19.5-4.4=15.1$ in units of $10^{-10}$
\cite{g-2SM2lA},  have to be taken into account (see
\citeres{g-2review2} for reviews of all contributions). 

It is an interesting question whether the observed deviation
(\ref{deviationfinal}) is due to supersymmetric effects. The
supersymmetric one-loop contribution \cite{g-2MSSMf1l} 
is approximately given by 
\BE
\amu^{\SU,{\rm 1L}} = 13 \times 10^{-10} 
  \frac{\tb\, {\rm  sign}(\mu)}{\left(\msusy/100\gev\right)^2} ,
\label{susy1loop}
\EE
if all supersymmetric particles (the relevant ones are the smuon,
sneutralino, chargino and neutralino) have a common mass
$\msusy$. 

This formula shows that supersymmetric
effects can easily account for a 
$(20\ldots30)\times10^{-10}$ deviation, if $\mu$ is positive and
$\msusy$ lies roughly between 100 GeV (for small $\tb$) and
600 GeV (for large $\tb$).
% This mass range is both allowed by
%present search limits and very interesting in view of physics at Run~II
%of the Tevatron, the LHC and a future Linear Collider (LC).
On the other hand, the precision of the
measurement places strong bounds on the supersymmetric parameter
space. 

%% In order to fully exploit the precision of the $(g-2)_\mu$ experiment 
%% within supersymmetry, the theoretical
%% uncertainty of the SUSY loop contributions from unknown higher-order
%% corrections should be significantly lower than the experimental error
%% and the hadronic uncertainties in the SM
%% prediction. Thus, the reduction of the uncertainty of the SUSY loop
%% contributions down to the level of about $\pm1\times10^{-10}$ is
%% desirable. This accuracy has not been
%% reached so far, since the SUSY two-loop contributions are
%% largely unknown. 

In this talk we present the results of \citeres{g-2FSf,g-2ChaNeu} for
the Minimal Supersymmetric Standard Model (MSSM) two-loop
contributions of diagrams involving either a sfermion or a
chargino/neutralino subloop.
These contributions constitute the class of
two-loop contributions to $\amu$, where a supersymmetric loop is
inserted into a SM (or more precisely a two-Higgs-doublet model)
one-loop diagram.  

%% In general, every diagram in the MSSM must contain one continuous line
%% carrying the $\mu$-lepton number. Thus, the MSSM diagrams can be
%% divided into two classes: (1) diagrams 
%% containing a one-loop diagram involving supersymmetric particles
%% (with a $\tilde\mu$ or
%% $\tilde\nu_\mu$ line) to which a second loop is attached, and
%% (2) diagrams where a second loop is attached to a two-Higgs-doublet
%% model one-loop diagram (with a $\mu$ or $\nu_\mu$ line).

%% Before the appearance of \citeres{g-2FSf,g-2ChaNeu}, only two
%% parts of the two-loop contributions have been evaluated: the leading
%% $\log \KL m_\mu/\msusy\KR$-terms of supersymmetric one-loop diagrams
%% with an additional photon loop \cite{g-2MSSMlog2l}, an important
%% contribution of class (1), and certain approximations of the two-loop
%% diagrams with sfermion loops \citeres{g-2BarrZee1,g-2BarrZee2}.

These diagrams are particularly interesting
since they can depend on other parameters than the supersymmetric
one-loop diagrams and can therefore change the qualitative behaviour
of the supersymmetric contribution to $\amu$. In particular, they could
even be large if the one-loop contribution is suppressed, e.g.\ due to
heavy smuons and sneutrinos.

%% The sfermion loop and the chargino/neutralino loop contributions
%% presented here complete (together with the pure two-Higgs-doublet
%% model contributions calculated in \citeres{g-2FSf,g-2ChaNeu}) the
%% evaluation of the second class of supersymmetric contributions. As we
%% will see below, they can substantially modify the supersymmetric
%% one-loop contributions and show an interesting dependence on the
%% supersymmetric parameters.

\section{Calculation}

The diagrams we have to calculate are the two-loop three-point
graphs for the $\mu\mu\ga$ interaction with a closed sfermion or
chargino/neutralino subloop (and the corresponding counterterm
diagrams). The main calculational steps are the following: The
amplitudes for $\amu$ are generated using the program
\fa~\cite{feynarts,fa-mssm}, and the appropriate
projector~\cite{g-2SM2lA} is applied. The Dirac algebra and
the conversion to a linear combination of \twol\ integrals is
performed using \tc~\cite{2lred}.

%%  In order to simplify the integrals,
%% a large mass expansion \cite{smirnov} is applied where the muon mass
%% is taken as small and all other masses as large. All resulting \twol\
%% integrals are either \twol\ vacuum integrals or products of \onel\
%% integrals. They can be reduced to standard integrals, the two-loop
%% vacuum master integral $T_{134}$ and the one-loop functions $A_0$ and
%% $B_0$, and can be evaluated analytically. 

The main part of the two-loop calculation consists of the calculation of
the Feynman integrals and the simplification of their
coefficients, both of which is complicated by the large number of
different mass scales and the involved structure of the MSSM Feynman
rules. 

As a first step we perform a large mass expansion \cite{smirnov} where
the muon mass is taken as small and all other masses as large. 
%the ratio $m_\mu/M_{\rm heavy}$ up to ${\cal   O}({m_\mu^2}/{M_{\rm
%    heavy}^2})$, where $M_{\rm heavy}$ stands for all 
%heavy masses, $M_{Z,W}$ and Higgs and chargino/neutralino masses. 
The large mass expansion separates the light and heavy scales in the
two-loop integrals, with the following possibilities:
\\
\underline{(light 0-loop)$\circ$(heavy 2-loop)}: The heavy
  two-loop integral has no external momentum, hence it can be reduced
  to the two-loop vacuum master integral.\\
\underline{(light 1-loop)$\circ$(heavy 1-loop)}: The integrals
  can be reduced to the standard one-loop functions $A_0(m)$
  and $B_0(m_\mu^2,0,m_\mu)$.\\
Since all integrals contain at least one sfermion or
chargino/neutralino loop, which is heavy, the third case of a light
2-loop integral does not appear. This situation changes in the
calculation of the pure two-Higgs-doublet model diagrams
\cite{g-2ChaNeu}.

A few remarks have to be made in order to justify our use of
dimensional regularization with anticommuting $\ga_5$. Firstly, one
might expect that supersymmetry-restoring counterterms are necessary
as discussed in \citeres{susyrestore}. This is however not the case
since in the present calculation only two-Higgs-doublet model
counterterms appear, which are not related by supersymmetry. Hence,
generating the counterterms by multiplicative renormalization is
sufficient.

Secondly, there is the well-known inconsistency of using anticommuting
$\ga_5$ together with the trace formula 
\BE{\rm Tr}(\ga_5
\ga^\mu\ga^\nu\ga^\rho\ga^\si)\propto \epsilon^{\mu\nu\rho\si},
\EE
which is necessary in order to reproduce the correct four-dimensional
limit of the trace. 

Another problem connected with the $\epsilon$-tensor is that it is a
purely four-dimensional object, and contractions such as
$\epsilon^{\mu\alpha\beta\ga}\epsilon^{\nu}_{\ \alpha\beta\ga}$ 
or $\epsilon^{\mu\nu\alpha\beta}\epsilon^{\rho\si}{}_{\alpha\beta}$
have to be evaluated in four dimensions. This seems to contradict the
application of the projection operator from \citeres{g-2SM2lA} to
extract $\amu$, which relies on a purely $D$-dimensional
covariant decomposition of the regularized $\mu\mu\ga$ vertex
function. 

In the discussion of both problems connected with $\ga_5$ and the
$\epsilon$-tensor there is one main point to be noticed. The
triangle subdiagrams with three external vector bosons and external
momentum $k$, where the traces and $\epsilon$-tensors originate, do
not lead to the covariant $\epsilon^{\mu\nu\rho\si}k_\si$ of
power-counting degree +1. This covariant is multiplied with the same
prefactor as the chiral gauge anomaly, which is zero after summation
over a full generation of fermions or all charginos/neutralinos. The
only remaining covariants containing the $\epsilon$-tensor have
power-counting degree $<0$. If this fact is used in the two-loop
diagrams with triangle subdiagram, one can show that the inconsistency
of the trace does not show up. Moreover, the covariants of the
$\mu\mu\ga$-vertex function involving $\epsilon$-tensors are finite.
Therefore the four-dimensional treatment of the $\epsilon$-tensors
does not spoil the validity of the projection operator.\footnote{
For more details see the discussion in \citere{g-2ChaNeu}. The
situation is similar but not identical to the one of $\mu$-decay
\cite{mudec}. In the latter case all external momenta can be set to
zero, whereas in the calculation of $\amu$ the external momentum of
the muon is non-zero, $p^2=m_\mu^2\ne0$. Hence there are more momenta
the $\epsilon$-tensor can be contracted with.}

\section{Results}

The results for the supersymmetric contributions to $\amu$ are not
numbers but functions of all the unknown MSSM parameters. However,
the parameter dependence and the corresponding phenomenological
discussion shows important differences between the sfermion
and chargino/neutralino loop contributions.

\begin{list}{---}{
\addtolength{\itemsep}{-1ex}
\addtolength{\topsep}{-1ex}
\addtolength{\leftmargin}{-1.5ex}}
\item
The sfermion loop contributions depend on the Higgs sector parameters
$\mu$ and $\tan\beta$ and the sfermion mass parameters in a rather
complicated way. Furthermore, it turns out that experimental
constraints on the MSSM parameter space significantly restrict the
possible sfermion loop contributions \cite{g-2FSf}. 
%For example, the Higgsino mass
%parameter $\mu$ appears in the Higgs--stop coupling and can thus
%enhance the contributions to $\amu$, but for light stop masses one can
%derive strong bounds on $\mu$ from the lower limit on the Higgs-boson
%mass. 
\item
In contrast, the chargino/neutralino loop contributions depend on
$\mu$, $\tan\beta$ and the gaugino mass parameter $M_2$ in a quite
straightforward way, and experimental constraints on the parameter
space have not much impact \cite{g-2ChaNeu}. 
%For example, the $\mu$-parameter acts as
%the Higgsino mass, and thus higher values of $\mu$ decrease the
%contributions to $\amu$.
\end{list}

\begin{figure}
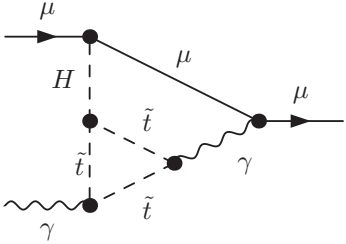

\vspace{-2em}
\begin{center}
\unitlength=1.0bp%
\begin{feynartspicture}(140,140)(1,1)

\FADiagram{}
\FAProp(0.,15.)(5.,15.)(0.,){/Straight}{1}
\FALabel(2.5,16.07)[b]{$\mu$}
\FAProp(0.,5.)(5.,5.)(0.,){/Sine}{0}
\FALabel(2.5,3.93)[t]{$\gamma$}
\FAProp(20.,10.)(15.,10.)(0.,){/Straight}{-1}
\FALabel(17.5,11.07)[b]{$\mu$}
\FAProp(5.,15.)(15.,10.)(0.,){/Straight}{0}
\FALabel(10.1014,13.1828)[bl]{$\mu$}
\FAProp(5.,10.)(5.,15.)(0.,){/ScalarDash}{0}
\FALabel(4.18,12.5)[r]{$H$}
{
\FAProp(5.,10.)(5.,5.)(0.,){/ScalarDash}{0}
\FALabel(4.18,7.5)[r]{$\tilde{t}$}
\FAProp(10.,7.5)(5.,10.)(0.,){/ScalarDash}{0}
\FALabel(7.60138,9.43276)[bl]{$\tilde{t}$}
\FAProp(10.,7.5)(5.,5.)(0.,){/ScalarDash}{0}
\FALabel(7.60138,5.56724)[tl]{$\tilde{t}$}}
\FAProp(10.,7.5)(15.,10.)(0.,){/Sine}{0}
\FALabel(12.7132,7.84364)[tl]{$\ga$}
\FAVert(10.,7.5){0}
{\FAVert(5.,10.){0}}
{\FAVert(5.,15.){0}}
\FAVert(5.,5.){0}
\FAVert(15.,10.){0}

\end{feynartspicture}
\end{center}
\vspace{-6em}
\caption{Diagram with a stop loop and photon and heavy Higgs-boson
  exchange. This diagram has a double enhancement and could give a
  large contribution to $\amu$.}
\label{fig:stoploop}
\end{figure}

In order to understand the parameter dependence of the sfermion loop
contributions in more detail, consider the diagram in
\reffi{fig:stoploop}. 
%This diagram has a stop loop, which couples to a
%photon and the heavy $\cp$-even Higgs boson $H$. 
It has a double
enhancement by the muon Yukawa coupling $\propto m_\mu \tan\beta$ and
by the Higgs--stop coupling $\propto \mu m_t$. Its value is
approximately given by \cite{g-2FSf}
\BEA
\label{stopcontrib}
\amu^{\Stop,{\rm 2L}} &\approx&-13\times10^{-10}\ {\rm sgn}(A_t)\times
\\&&
\left(\frac{\tan\beta}{50}\right)
{\left(\frac{\mu}{20\ m_{\tilde{t}}}\right)}\left(\frac{m_t}{M_H}\right),
\nonumber
\EEA
provided that $m_{\Stop}\lsim M_H$ ($\Stop$ stands for lightest stop
mass eigenstate). For large
$\tan\beta\approx50$ and $M_H$ around $200\gev$, we find that the stop
diagram contribution is mainly determined by the ratio
$\mu/m_{\Stop}$. If this ratio is very large, $\mu/m_{\Stop}\gsim20$,
contributions of more than $10\times10^{-10}$ are possible. This was
already noticed in \citere{g-2BarrZee2} and would correspond to
a two-loop contribution of more than $1\ldots2$ experimental standard
deviations. 

Whether or not the ratio $\mu/m_{\Stop}$ is restricted by experimental
constraints depends on yet another property of the MSSM parameters,
namely the universality between the stop and sbottom mass
parameters. The comparison of the experimental limit and the
MSSM-prediction of the lightest Higgs-boson mass $M_h$
\cite{mhiggsletter,feynhiggs} generally constrains the ratio
(off-diagonal):(diagonal entry) in the stop mass matrix.
If the stop and sbottom mass parameters are universal, this translates
into a limit on the ratio $\mu/m_{\Stop}\lsim3$ for the relevant
parameter space. As the ratio between sbottom and stop mass parameters
increases, also larger values for $\mu/m_{\Stop}$ become possible.

%% However, the ratio $\mu/m_{\Stop}$ can be severely constrained by the
%% experimental limit on the lightest Higgs-boson mass. From comparing
%% this limit with the MSSM prediction of 
%% $M_h$ in terms of the MSSM parameters \cite{mhiggsletter} one obtains
%% the restriction $X_t/\msusy\lsim2.5$ between the
%% stop-mixing parameter $X_t$ and the diagonal stop mass parameter
%% $\msusy$. If universality between the stop and the sbottom mass
%% parameters is assumed, this restriction translates into the
%% approximate relation $\mu/\msusy2.5$ for large $\tan\beta$.
%% Since the stop mass eigenvalues are mainly determined by $\msusy$ this
%% entails a restriction on the ratio $\mu/m_{\Stop}$.

%%%%%%%%%%%%% F I G U R E %%%%%%%%%%%%%%%%%%%%%%%%%%%%%%%%%%%%
\begin{figure}
\BC
\epsfig{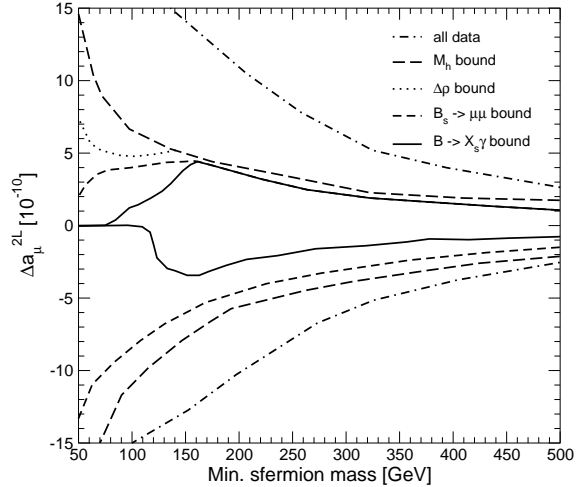}
\vspace{-4em}
\caption{%
Maximum contributions of the two-loop diagrams with a closed sfermion loop 
to $\amu$ as a function of the lightest sfermion mass. No constraints on the
MSSM parameters are taken into account for the outermost curve. Going
to the inner curves additional constraints (see text) have
been applied.
}
\label{fig:gammaWscan}
\EC
\vspace{-2em}
\end{figure}

In \reffi{fig:gammaWscan} we show the full results for the sfermion
contributions as functions of the lightest sfermion mass for universal
sfermion mass parameters. The outer
lines show the maximum possible results for $\tan\beta=50$ if all
experimental constraints are ignored and all MSSM
mass parameters are varied universally up to $3\tev$ (for the $\cp$-odd
Higgs-boson mass we use $M_A>150\gev$). The next lines show the
maximum possible results if the aforementioned experimental limit on
$M_h$ is taken into account. We find, in agreement with the preceding
discussion, that the maximum results are drastically reduced. For a
lightest sfermion mass of $100\gev$, the results are reduced from more
than $15\times10^{-10}$ to about $5\times10^{-10}$. The inner lines
correspond to taking into account more experimental constraints on
$\De\rho$ \cite{delrhosusy2loop}, $\br(B_s\to\mu^+\mu^-)$~\cite{bsmumu} and 
  $\br(B \to X_s\ga)$~\cite{bsg}. They reduce the maximum
contributions further.

%% From the above discussion we can see that non-universal sfermion mass
%% parameters could make a larger ratio $\mu/m_{\Stop}$ compatible with
%% the limit on $M_h$. As shown in \citere{g-2FSf}, contributions to
%% $\amu$ of more than $15\times10^{-10}$ are indeed possible if the
%% ratio between the sbottom and the stop mass parameters is of order 10
%% or more. 

The chargino/neutralino two-loop contributions have a more straightforward
parameter dependence. They depend on $\tan\beta$ and the mass
parameters for the Higgsinos, $\mu$, the gauginos, $M_2$, and the
$\cp$-odd Higgs boson, $M_A$. For the simple case that all these mass
parameters are equal to a common mass scale $\msusy$, we obtain the
approximation
\BE
\amu^{\chi,\rm 2L} = 11 \times 10^{-10} 
  \frac{(\tb/50)\, {\rm  sign}(\mu)}{\left(\msusy/100\gev\right)^2} .
\label{chaneuapprox}
\EE
If all the masses are even equal to the smuon and sneutrino masses,
this formula can be immediately compared to the one-loop contributions
(\ref{susy1loop}). In this case the chargino/neutralino two-loop
contributions amount to about 2\%\ of the one-loop contributions.

\begin{figure}[htb]
\vspace{-2em}
\unitlength=21.0bp%
\begin{picture}(10,10)\epsfxsize=7.5cm
                  \put(0,0){\epsfbox{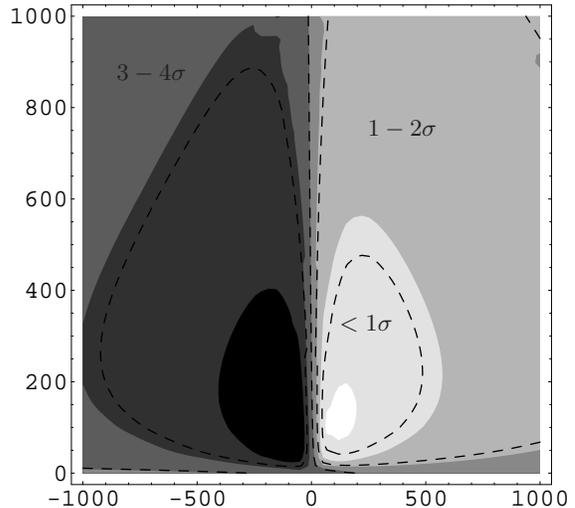}}
\put(6,3.5){\small$<1\sigma$}
\put(2,8){\small$3-4\sigma$}
\put(6.5,7){\small$1-2\sigma$}
\end{picture}
\vspace{-2em}
\caption{Contour plot of $\amu^{\SU,\rm 1L}+\amu^{\chi,\rm
  2L}$ (fully drawn areas) and $\amu^{\SU,\rm 1L}$ (dashed contours)
  in the $\mu$--$M_2$-plane. The borders of the regions and the
  contours correspond to $1\si$, $2\si$, \ldots deviation from the
  observed value according to eq.\ (\ref{deviationfinal}).}
\vspace{-2em}
\label{fig:chaneuplot}
\end{figure}

If the smuon and sneutrino masses are heavier than the chargino and
neutralino masses, the one-loop contributions are suppressed and the
two-loop contributions can have a larger
impact. Fig.~\ref{fig:chaneuplot} shows the sum $\amu^{\SU,\rm
  1L}+\amu^{\chi,\rm 
  2L}$ in comparison to the one-loop result $\amu^{\SU,\rm 1L}$ alone
as a contour plot in the $\mu$--$M_2$-plane. The smuon and sneutrino
masses are fixed to $1\tev$ and $\tb=50$, $M_A=200$. We find that in
this case the two-loop corrections from the chargino/neutralino loop
diagrams can modify the $1\si$, $2\si$, \ldots contours significantly.

\section{Conclusions}

Supersymmetric contributions to $\amu$ could easily account for the
observed $(20\ldots30)\times10^{-10}$ deviation between SM theory and
experiment. Conversely, the precision of the experiment places
stringent bounds on the MSSM parameter space.

The two-loop contributions presented here can substantially modify the
supersymmetric one-loop contribution, and their knowledge reduces the
theoretical uncertainty of the supersymmetric prediction for
$\amu$. The sfermion loop contributions can have values up to
$5\times10^{-10}$ for small sfermion masses and large $\tb$; for
non-universal stop and sbottom mass parameters with a large hierarchy
they can be even larger. In the numerical evaluation of these
contributions it is crucial to take into account all known
experimental constraints on the MSSM parameter space.
The chargino/neutralino loop contributions are well approximated by
eq.\ (\ref{chaneuapprox}) and can have values of more
than $5\times10^{-10}$, if the chargino/neutralino masses are small. 

In \citeres{g-2FSf,g-2ChaNeu} also the SM/two-Higgs-doublet model-like
contributions of two-loop diagrams with fermion loops and with purely
bosonic loops have been computed. The difference of the diagrams in
the MSSM and the SM is smaller than $1\times10^{-10}$. In order to
complete the full two-loop calculation of $\amu$ in the MSSM, the
two-loop corrections to the supersymmetric one-loop diagrams with
smuon or sneutrino exchange have to be evaluated.

\paragraph{Acknowledgments}

I thank the organizers and the participants of Loops \& Legs 2004 for
creating a pleasant and stimulating atmosphere.


\begin{thebibliography}{99} 

\bibitem{g-2exp} [The Muon g-2 Collaboration],
%                 {\em Phys. Rev. Lett.} {\bf 89} (2002) 101804
%                 [Erratum-ibid.\  {\bf 89} (2002) 129903], 
%                 hep-ex/0208001; 
                 %%CITATION = HEP-EX 0208001;%%
%                 hep-ex/0301003;
                 %%CITATION = HEP-EX 0301003;%%
                 {\em Phys. Rev. Lett.} {\bf 92} (2004) 161802.
                 %%CITATION = HEP-EX 0401008;%%

\bibitem{DEHZ} M.~Davier, S.~Eidelman, A.~H\"ocker and Z.~Zhang,
               {\em Eur.\ Phys.\ J.}\  {\bf C 31} (2003) 503.
               %%CITATION = HEP-PH 0308213;%%

\bibitem{g-2HMNT} K.~Hagiwara, A.~Martin, D.~Nomura and T.~Teubner,
%                  {\em Phys. Lett.} {\bf B 557} (2003) 69, 
%                  hep-ph/0209187v2;
                  %%CITATION = HEP-PH 0209187;%%
                  {\em Phys.\ Rev.}\  {\bf D 69} (2004) 093003.
                  %%CITATION = HEP-PH 0312250;%%



\bibitem{Jegerlehner} S.~Ghozzi and F.~Jegerlehner,
                      {\em Phys. Lett.} {\bf B 583} (2004) 222.
                      %%CITATION = HEP-PH 0310181;%%

\bibitem{Yndurain} J.~de Troconiz and F.~Yndurain,
                   hep-ph/0402285.
                   %%CITATION = HEP-PH 0402285;%%

\bibitem{LBL} M.~Knecht and A.~Nyffeler,
              {\em Phys. Rev.} {\bf D 65} (2002) 073034;
              %%CITATION = HEP-PH 0111058;%%
              M.~Knecht, A.~Nyffeler, M.~Perrottet and E.~De Rafael,
              {\em Phys. Rev. Lett.} {\bf 88} (2002) 071802; 
              %%CITATION = HEP-PH 0111059;%%
              I.~Blokland, A.~Czarnecki and K.~Melnikov,
              {\em Phys. Rev. Lett.} {\bf 88} (2002) 071803;
              %%CITATION = HEP-PH 0112117;%%
              M.~Ramsey-Musolf and M.~Wise,
              {\em Phys. Rev. Lett.} {\bf 89} (2002) 041601;
              %%CITATION = HEP-PH 0201297;%%
              J.~Kuhn, A.~Onishchenko, A.~Pivovarov and O.~Veretin,
              {\em Phys. Rev.} {\bf D 68} (2003) 033018. 
              %%CITATION = HEP-PH 0301151;%%

\bibitem{LBLnew} K.~Melnikov and A.~Vainshtein,
                 hep-ph/0312226.
                 %%CITATION = HEP-PH 0312226;%%
\bibitem{g-2SM2lA} A.~Czarnecki, B.~Krause and W.~Marciano,
                   {\em Phys. Rev. Lett.}  {\bf 76} (1996) 3267;
                   %%CITATION = HEP-PH 9512369;%%
                   {\em Phys. Rev.} {\bf D 52} (1995) 2619;
                   %%CITATION = HEP-PH 9506256;%%
                   B.~Krause, 
                   PhD thesis, Universit\"at Karlsruhe, 1997,
                   Shaker Verlag, ISBN~3-8265-2780-1.

\bibitem{g-2review2} A.~Czarnecki and W.~Marciano,
                    {\em Phys. Rev.} {\bf D 64} (2001) 013014;
                    %%CITATION = HEP-PH 0102122;%%
                     M.~Knecht,
                     hep-ph/0307239.
                     %%CITATION = HEP-PH 0307239;%%

\bibitem{g-2MSSMf1l} T.~Moroi,
                     {\em Phys. Rev.} {\bf D 53} (1996) 6565
                     [Erratum-ibid.\ {\bf D 56} (1997) 4424].
                     %%CITATION = HEP-PH 9512396;%%

\bibitem{g-2FSf} S.~Heinemeyer, D.~St\"ockinger and G.~Weiglein,
                 {\em Nucl. Phys.} {\bf B 690} (2004) 62.
                 %%CITATION = HEP-PH 0312264;%%

\bibitem{g-2ChaNeu}
                 S.~Heinemeyer, D.~St\"ockinger and G.~Weiglein,
                 hep-ph/0405255.
                 %%CITATION = HEP-PH 0405255;%%


%\bibitem{g-2MSSMlog2l} G.~Degrassi and G.~Giudice,
%                       {\em Phys. Rev.} {\bf D 58} (1998) 053007, 
%                       hep-ph/9803384.
                       %%CITATION = HEP-PH 9803384;%%

%\bibitem{g-2BarrZee1} C.~Chen and C.~Geng,
%                      {\em Phys. Lett.} {\bf B 511} (2001) 77, 
%                      hep-ph/0104151.
                      %%CITATION = HEP-PH 0104151;%%

\bibitem{feynarts} J.~K\"ublbeck, M.~B\"ohm, and A.~Denner,
                   {\em Comput. Phys. Commun.} {\bf 60} (1990) 165; 
                   %%CITATION = CPHCB,60,165;%%
                   T.~Hahn,
                   {\em Comput. Phys. Commun.} {\bf 140} (2001) 418.
                   %%CITATION = HEP-PH 0012260;%%

\bibitem{fa-mssm} T.~Hahn and C.~Schappacher,
                  {\em Comput. Phys. Commun.} {\bf 143} (2002) 54.
                  %%CITATION = HEP-PH 0105349;%%

\bibitem{2lred} G.~Weiglein, R.~Scharf and M.~B\"ohm,
                {\em Nucl. Phys.} {\bf B 416} (1994) 606.
                %%CITATION = HEP-PH 9310358;%%
%                G.~Weiglein, R.~Mertig, R.~Scharf and M.~B\"ohm, 
%                in {\it New Computing Techniques in Physics Research 2},
%                ed.~D.~Perret-Gallix (World Scientific, Singapore,
%                1992), p.~617.

\bibitem{smirnov} V.~Smirnov, {\em Applied Asymptotic Expansions in
                Momenta and Masses}, Springer Verlag, Berlin (2002).


%\bibitem{t134} A. Davydychev und J.~Tausk, 
%               {\em Nucl. Phys.} {\bf B 397} (1993) 123;\\
%               %%CITATION = NUPHA,B397,123;%%
%               F. Berends und J.~Tausk,
%               {\em Nucl. Phys.} {\bf B 421} (1994) 456.
%               %%CITATION = NUPHA,B421,456;%%
%\bibitem{a0b0c0d0} G.~Passarino and  M.~Veltman, 
%                   {\em Nucl. Phys.} {\bf B 160} (1979) 151.
%                   %%CITATION = NUPHA,B153,365;%%

\bibitem{susyrestore} W.~Beenakker, R.~H\"opker and P.~M.~Zerwas,
                      {\em Phys.\ Lett.\ } {\bf B 378} (1996) 159.
                      %%CITATION = HEP-PH 9602378;%%
                      W.~Hollik, E.~Kraus and D.~St\"ockinger,
                      {\em Eur. Phys. J.} {\bf C 11} (1999) 365;
                      %%CITATION = HEP-PH 9907393;%%
                      W.~Hollik and D.~St\"ockinger,
                      {\em Eur. Phys. J.} {\bf C 20} (2001) 105;
                      %%CITATION = HEP-PH 0103009;%%
           W.~Hollik, E.~Kraus, M.~Roth, C.~Rupp, K.~Sibold and D.~St\"ockinger,
           {\em Nucl. Phys.} {\bf B 639} (2002) 3;
           %%CITATION = HEP-PH 0204350;%%
           I.~Fischer, W.~Hollik, M.~Roth and D.~St\"ockinger,
           {\em Phys. Rev.} {\bf D 69} (2004) 015004.

\bibitem{mudec} A.~Freitas, W.~Hollik, W.~Walter and G.~Weiglein,
           %``Complete fermionic two-loop results for the M(W)-M(Z)
           %interdependence,''
           {\em Phys.\ Lett.} {\bf B 495} (2000) 338
           [Erratum-ibid.\ {\bf B 570} (2003) 260];
           %%CITATION = HEP-PH 0007091;%%
%           A.~Freitas, W.~Hollik, W.~Walter and G.~Weiglein,
    %``Electroweak two-loop corrections to the M(W) - M(Z) mass correlation
    %in  the standard model,''
           {\em Nucl.\ Phys.} {\bf B 632} (2002) 189
           [Erratum-ibid.\ {\bf B 666} (2003) 305].
           %%CITATION = HEP-PH 0202131;%%


\bibitem{g-2BarrZee2} A.~Arhrib and S.~Baek,
                      {\em Phys. Rev.} {\bf D 65} (2002) 075002.
                      %%CITATION = HEP-PH 0104225;%%
\bibitem{mhiggsletter}G.~Degrassi, S.~Heinemeyer, W.~Hollik,
                     P.~Slavich and G.~Weiglein, 
                     {\em Eur. Phys. Jour.} {\bf C 28} (2003) 133;
                     %%CITATION = HEP-PH 0212020;%%
                     for further references see S. Heinemeyer, these
                     proceedings; hep-ph/0406245.
%%CITATION = HEP-PH 0406245;%% 
%%                       S.~Heinemeyer, W.~Hollik and G.~Weiglein, 
%%                        {\em Phys. Rev.} {\bf D 58} (1998) 091701;
%%                        %%CITATION = HEP-PH 9803277;%%
%%                        {\em Phys. Lett.} {\bf B 440} (1998) 296;
%%                        %%CITATION = HEP-PH 9807423;%%
%%  %                     S.~Heinemeyer, W.~Hollik and G.~Weiglein, 
%%                      {\em Eur. Phys. Jour.} {\bf C 9} (1999) 343;
%%                      %%CITATION = HEP-PH 9812472;%%
%%                      G.~Degrassi, S.~Heinemeyer, W.~Hollik,
%%                     P.~Slavich and G.~Weiglein, 
%%                     {\em Eur. Phys. Jour.} {\bf C 28} (2003) 133;
%%                     %%CITATION = HEP-PH 0212020;%%
%%                     A.~Dedes, G.~Degrassi and P.~Slavich,
%%                     {\em Nucl.\ Phys.}\  {\bf B 672} (2003) 144.
%%                     %%CITATION = HEP-PH 0305127;%%
\bibitem{feynhiggs} S.~Heinemeyer, W.~Hollik, G.~Weiglein,
                    {\em Comp. Phys. Comm.} {\bf 124} 2000 76;
%                    hep-ph/9812320;
                    %%CITATION = HEP-PH 9812320;%%
                    see {\tt www.feynhiggs.de} .

\bibitem{delrhosusy2loop} For references see S. Heinemeyer, these proceedings.
%%                           A.~Djouadi, P.~Gambino, S.~Heinemeyer, W.~Hollik,
%%                           C.~J\"unger and G.~Weiglein,
%%                           {\em Phys. Rev. Lett.} {\bf 78} (1997) 3626;
%%                           %%CITATION = HEP-PH 9612363;%%
%%                           {\em Phys. Rev.} {\bf D 57} (1998) 4179;
%%                           %%CITATION = HEP-PH 9710438;%%
%%                           S.~Heinemeyer and G.~Weiglein, 
%%                           {\em JHEP} {\bf 0210} (2002) 072;
%%                           %%CITATION = HEP-PH 0209305;%%
%%                           hep-ph/0301062.
%%                           %%CITATION = HEP-PH 0301062;%%
                      
\bibitem{bsmumu}
%%  F.~Azfar, 
%%                  hep-ex/0309005;\\
%%                  %%CITATION = HEP-EX 0909005;%%
%% % , presented at {\em``XXIII
%% %   Physics in Collision'', Zeuthen, Germany, 26--28 June 2003}.
%%                  M.~Nakao,
%%                  talk given at {\em LeptonPhoton 2003}, Fermilab,
%%                  August 2003;\\
%%                  %\cite{Babu:1999hn}
%% K.~Babu and C.~Kolda,
%% {\em Phys.\ Rev.\ Lett.}\  {\bf 84} (2000) 228,
%% hep-ph/9909476;\\
%% %%CITATION = HEP-PH 9909476;%%
%% S.~Choudhury and N.~Gaur,
%% %``Dileptonic decay of B/s meson in SUSY models with large tan(beta),''
%% {\em Phys.\ Lett.}\  {\bf B 451} (1999) 86,
%% hep-ph/9810307;\\
%% %%CITATION = HEP-PH 9810307;%%
%% C.~Bobeth, T.~Ewerth, F.~Kruger and J.~Urban,
%% {\em Phys.\ Rev.}\  {\bf D 64} (2001) 074014,
%% hep-ph/0104284;\\
%% %%CITATION = HEP-PH 0104284;%%
%% A.~Dedes, H.~Dreiner and U.~Nierste,
%% {\em Phys.\ Rev.\ Lett.}\  {\bf 87} (2001) 251804,
%% hep-ph/0108037;\\
%% %%CITATION = HEP-PH 0108037;%%
%% G.~Isidori and A.~Retico,
%% {\em JHEP} {\bf 0111} (2001) 001,
%% hep-ph/0110121;\\
%% %%CITATION = HEP-PH 0110121;%%
%% A.~Dedes and A.~Pilaftsis,
%% {\em Phys.\ Rev.}\  {\bf D 67} (2003) 015012,
%% hep-ph/0209306;\\
%% %%CITATION = HEP-PH 0209306;%%
%% A.~Buras, P.~Chankowski, J.~Rosiek and L.~Slawianowska,
%% {\em Nucl.\ Phys.}\  {\bf B 659} (2003) 3,
%% hep-ph/0210145;\\
%%CITATION = HEP-PH 0210145;%%
A.~Dedes,
{\em Mod.\ Phys.\ Lett.}\  {\bf A 18} (2003) 2627,
hep-ph/0309233 and references therein.
%%CITATION = HEP-PH 0309233;%%

\bibitem{bsg} T.~Hurth,
%``Present status of inclusive rare B decays,''
Rev.\ Mod.\ Phys.\  {\bf 75} (2003) 1159 and references therein.
%%CITATION = HEP-PH 0212304;%%

\end{thebibliography}
\end{document}